\documentclass{mem}

\usepackage{natbib}
\usepackage{graphicx}
\usepackage{txfonts}

\usepackage{hyperref}
\usepackage{color}

\bibpunct{(}{)}{;}{a}{}{,}
\newcommand{\eprint}[1]{\url{#1}}

\begin{document}
\title{Simulations of Electron Capture and Low-Mass Iron Core Supernoave}
\titlerunning{Electron Capture and Low-Mass Iron Core Supernovae}
\authorrunning{B.~M\"uller et al.}
\author{B.~M\"uller\inst{1,2}, S.~Wanajo\inst{3,4}, H.-Th.~Janka\inst{5}, A.~Heger\inst{2},
D.~Gay\inst{1}, S.A.~Sim\inst{1}
}
\institute{Astrophysics Research Centre, School of Mathematics
  and Physics, Queen’s University Belfast, Belfast, BT7 1NN, United
  Kingdom\\
  \email{b.mueller@qub.ac.uk}
  \and
  Monash Centre for Astrophysics, School of Physics and Astronomy, Monash University, VIC 3800, Australia
  \and
  Department of Engineering and Applied Sciences,
  Sophia University, Chiyoda-ku, Tokyo 102-8554, Japan
  \and
  iTHES Research Group, RIKEN, Wako, Saitama 351-0198, Japan
  \and
  Max-Planck-Institut f\"ur Astrophysik,
  Karl-Schwarzschild-Str. 1, 85748 Garching, Germany
}

\abstract{The evolutionary pathways of core-collapse supernova
  progenitors at the low-mass end of the spectrum are beset with major
  uncertainties. In recent years, a variety of evolutionary channels
  has been discovered in addition to the classical electron capure
  supernova channel of super-AGB stars.  The few available progenitor
  models at the low-mass end have been studied with great success in
  supernova simulations as the peculiar density structure
  makes for robust neutrino-driven explosions in this mass
  range. Detailed nucleosynthesis calculations have been conducted
  both for models of electron capture supernovae and low-mass iron
  core supernovae and revealed an interesting production of the
  lighter trans-iron elements (such as Zn, Sr, Y, Zr) as well as 
  rare isotopes
like ${}^{48}\mathrm{Ca}$
and ${}^{60}\mathrm{Fe}$. We stress the need to explore the
low-mass end of the supernova spectrum further and link various
observables to understand the diversity of explosions in this regime.}

\maketitle

\section{Introduction}

The region just above the minimum mass for core-collapse supernova (SN)
explosions is of particular interest for supernova theory for several
reasons. Due to the steepness of the initial mass function, roughly
$20 \%$ of core-collapse events originate from progenitors
within $2 M_\odot$ of this minimum mass
on the zero-age main sequence.

\begin{figure}
  \includegraphics[width=\linewidth]{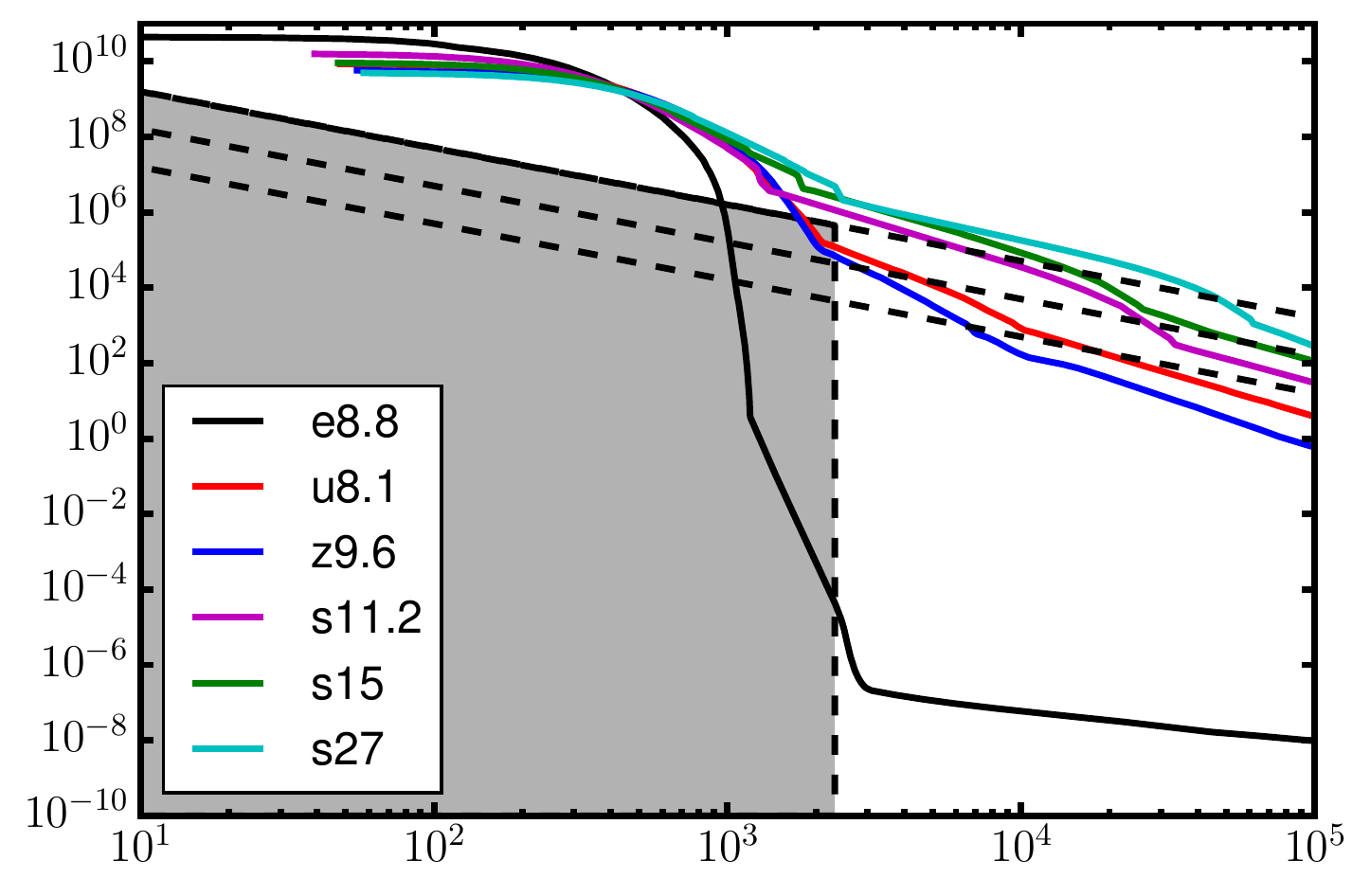}
  \caption{Density profiles of an ECSN
progenitor (e8.8, \citealp{nomoto_84}),
low-mass iron core progenitors
with $8.1 M_\odot$ (u8.1, $Z=10^{-4}Z_\odot$)
and $9.6 M_\odot$ (z9.6, $Z=0$), and more massive
progenitors with $11.2 M_\odot$, $27 M_\odot$
(s11.2 and s27, \citealp{woosley_02}) and $15 M_\odot$
\citep{woosley_95}. The approximate accretion
rate is indicated by
slanted dashed lines ($0.05 M_\odot \, \mathrm{s}^{-1}$,
$0.005 M_\odot \, \mathrm{s}^{-1}$, and
$5\times 10^{-4} M_\odot \, \mathrm{s}^{-1}$
from the top).
The vertical dashed line roughly
indicates an infall time
of $0.5 \, \mathrm{s}$. ECSN-like explosions
with fast shock expansion and
without significant accretion after
shock revival are expected
in the grey-shaded region (see \citealt{mueller_16b}
for details). Note that
the low-mass iron core progenitors only
marginally fall into this regime and that
the transition from the ``ECSN-like'' regime
to normal supernovae is not abrupt in reality.
\label{fig:profiles}}
\end{figure}

From the point of view of stellar evolution, the lower end of the mass
range for core-collapse SNe is remarkably different from
generic high-mass progenitors. Contrary to higher masses, degeneracy
and off-centre ignition play a major role during the evolution beyond
carbon burning, which lead to structural peculiarities of the
progenitors. The classical ``electron capture supernova'' (ECSN)
channel for super-AGB progenitors (SAGB) best exemplifies these
peculiarities \citep{nomoto_84,nomoto_87}. Here collapse is triggered by
electron captures on ${}^{20}\mathrm{Ne}$ and ${}^{24}\mathrm{Mg}$ in
a degenerate O-Ne-Mg core, which is separated merely by a tiny C/O
layer from the H envelope as the He shell has been eliminated by
dredge-up.  While the width of the classical ECSN
channel is beset with uncertainties such as our incomplete
understanding of O ignition and flame propagation \citep{timmes_92,timmes_94,jones_16},
studies of the AGB-SN mass transition in recent years
\citep{jones_13,jones_14,doherty_15,doherty_17,woosley_15} have
unearthed a variety of pathways towards collapse that lead to a similar
-- though sometimes less extreme -- progenitor structure characterised
by a strongly degenerate core with a very steep density gradient into
the surrounding tenuous shells.

While the intricacies of stellar evolution at the AGB-SN mass
transition still present a challenge, this mass range has been a
particularly fruitful target for first-principle supernova simulations
for the last decade since the first modern ECSN
explosion model of \citet{kitaura_06}. Contrary to more
massive progenitors, the explosion mechanism close to this transition
is understood to the degree that neutrino-driven explosions are readily
obtained in self-consistent simulations in 1D
\citep{kitaura_06,fischer_10,huedepohl_10,melson_15a,radice_17}, 2D
\citep{wanajo_11,mueller_12b,janka_12b,radice_17}, and 3D \citep{melson_15a}.

\section{Explosion Dynamics of Electron Capture and Low Mass Iron Core Supernovae}
The critical structural feature behind the robustness
of neutrino-driven shock revival close to the AGB-SN mass
transition is the steep density gradient outside the core
(Fig.~\ref{fig:profiles}).
This results in a rapid drop of the mass accretion 
rate $\dot{M}$ onto the proto-neutron star 
(of mass $M$) soon after bounce,
which is related to the initial density $\rho$ of infalling shells
from radius $r$ as
$\dot{M} \approx 8 \pi \rho \sqrt{Gm r^3/3}$.
Consequently, the
stagnation radius of the shock
increases due to the lower pre-shock ram pressure, conditions become
favourable to neutrino-driven runaway shock expansion in low-mass SN
progenitors early on.

As $\dot{M}$ plummets rapidly, these explosions do
not exhibit an extended phase of concurrent accretion and mass
ejection after shock revival, which can last for seconds in more
massive progenitors \citep{mueller_15b,bruenn_16,mueller_17}.
Without such a cycle of accretion and mass ejection, the explosion
energy is essentially set by the mass in the gain region around the
onset of shock revival.  Once the ejected matter is unbound by
neutrino heating, the residual net contribution to the explosion
energy is provided by the recombination of nucleons into heavy nuclei
and $\alpha$-particles \citep{janka_08}.  The ejection of
$\mathord{\sim}0.01 M_\odot$ results in a small explosion energy of
$\mathord{\sim} (0.5 \ldots 1) \times 10^{50} \, \mathrm{erg}$ with
a small additional contribution from the neutrino-driven wind
on longer time-scales. The explosion dynamics allows only for the production of a small
amount of radioactive ${}^{56}\mathrm{Ni}$ of the order
of $10^{-3} M_\odot$.

Although shock revival is found even in 1D models
of progenitors close to the AGB-SN transition, multi-D effects are
not negligible for the explosion dynamics, especially for models with
less extreme density profiles. While multi-D effects only
boost the explosion energy by $\mathord{\sim}{10\%}$ in models of the
  classical ECSN channel, they do enhance the
  explosion energy by a factor of several for low-mass iron core
  progenitors with more substantial C/O and He shells
  \citep{melson_15a,radice_17}.
Nevertheless, rapid shock revival and the absence of
an extended phase of concurrent accretion and mass
ejection motivate  classifying such explosions
as ``ECSN-like'' as opposed to SNe from
more massive progenitors.

\begin{figure*}
  \includegraphics[width=0.49\linewidth]{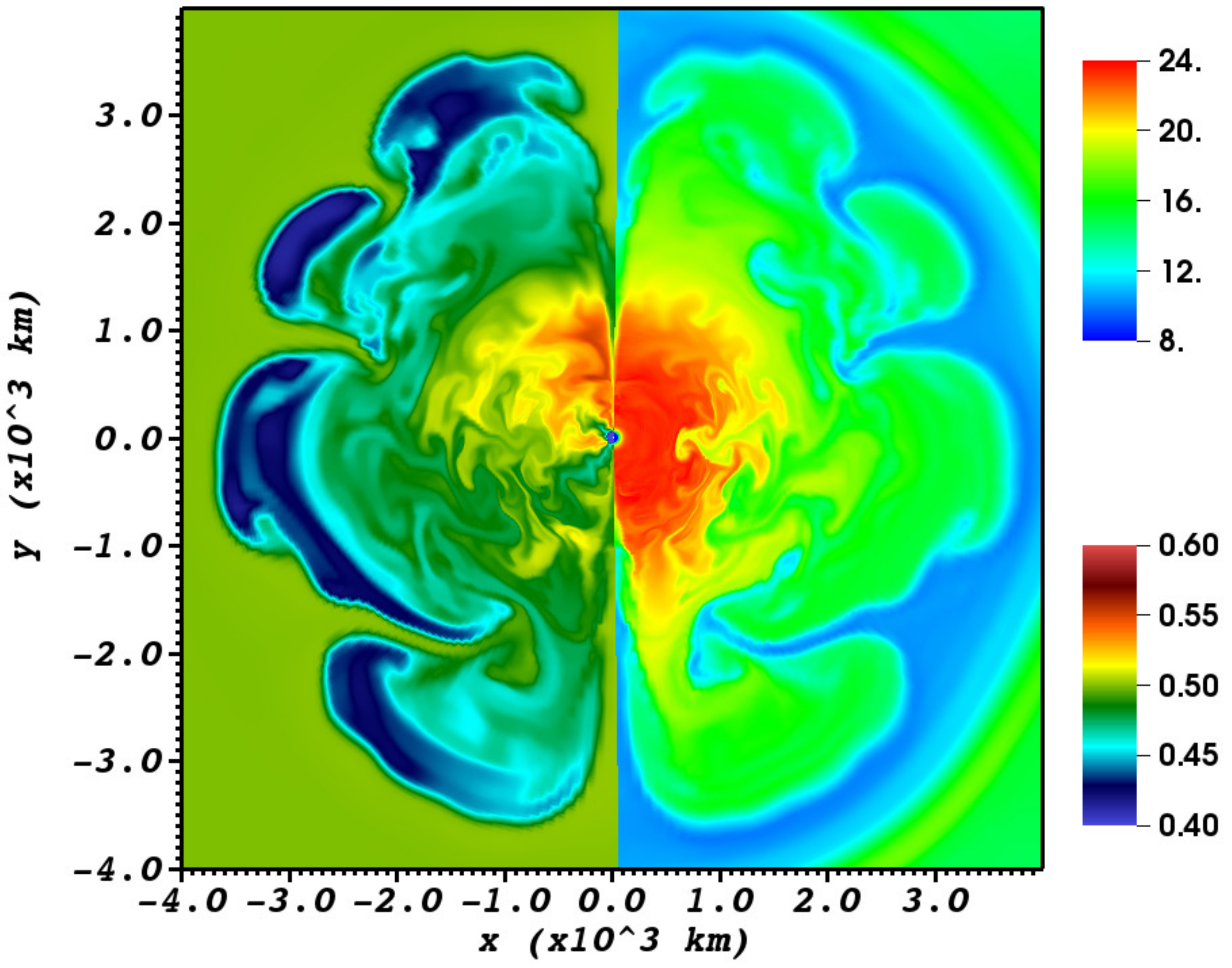}
  \includegraphics[width=0.49\linewidth]{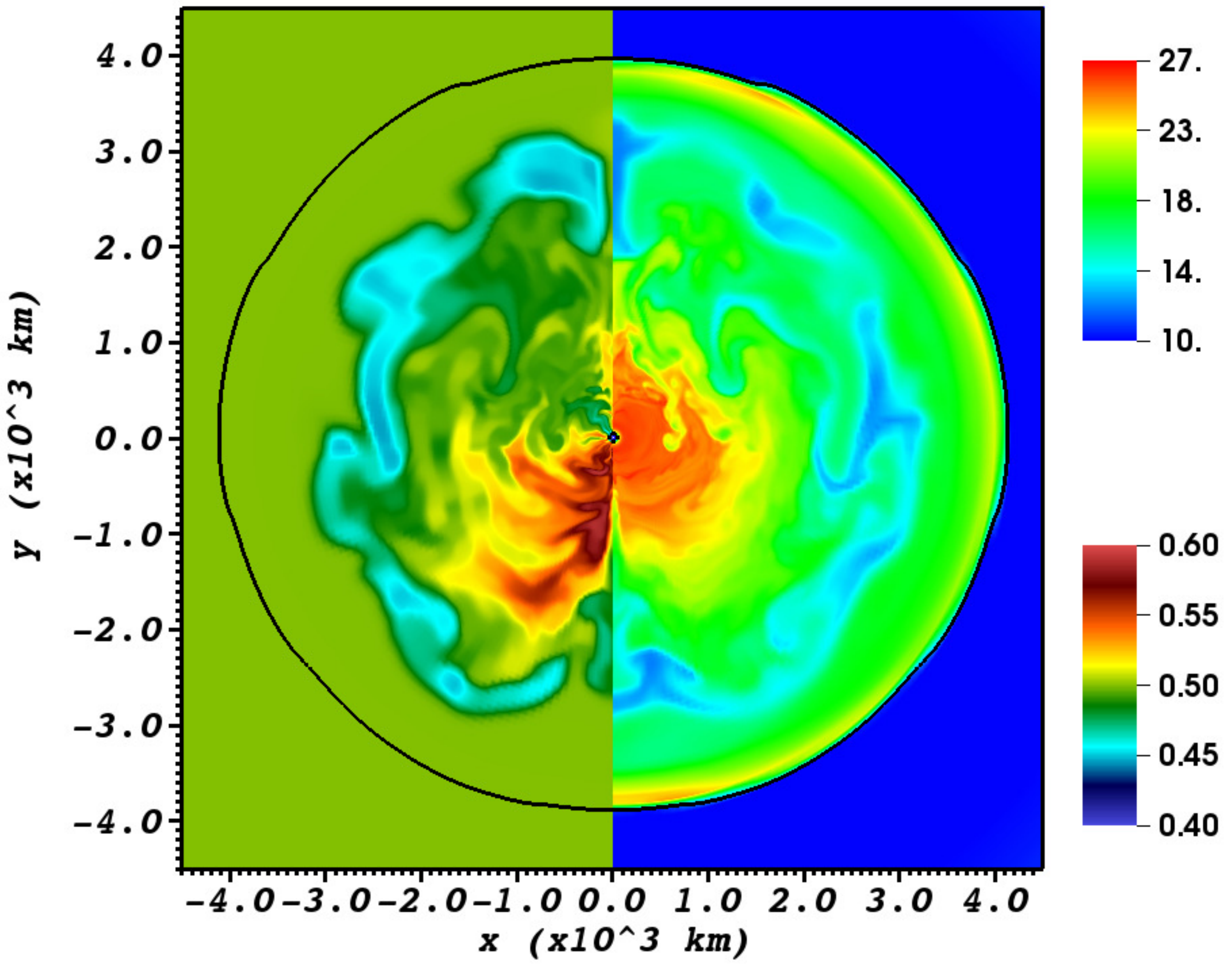}
  \caption{Electron fraction $Y_e$ (left half of panels)
and entropy $s$ (right half of panels) in 2D simulations
of an ECSN (left, $266\,                   
\mathrm{ms}$ after bounce, progenitor e8.8)
and in a low-mass iron core supernova
(right, $317\, \mathrm{ms}$ after bounce, progenitor z9.6).
Both show similar neutron-rich  Rayleigh-Taylor plumes
that develop shortly after shock revival.
\label{fig:2d}}
\end{figure*}

\section{Nucleosynthesis in the Neutrino-Processed Ejecta}
\begin{figure*}
  \includegraphics[width=\linewidth]{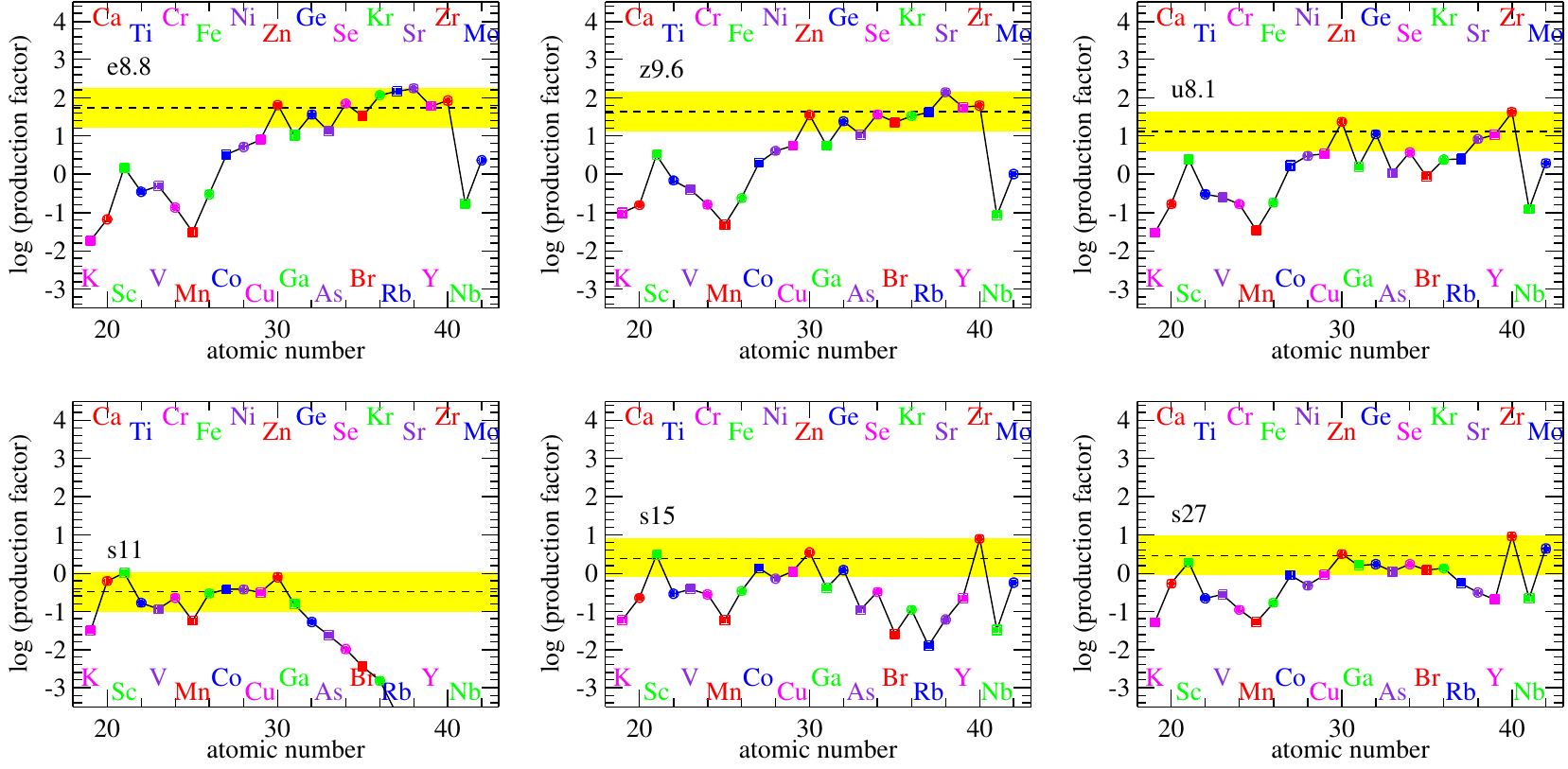}
  \caption{Elemental production factors for the ECSN model e8.8 and
    the low-mass iron core models z9.6 and u8.1 (top row) compared to
    the production factors for the more massive progenitor s11.2, s15,
    and s27 (bottom row), taken from \citet{wanajo_17}. The production factors are defined as the
    ratio of mass fraction of an element in the ejecta and the
    corresponding solar value \citep{lodders_03}. The yellow bands
    denote a range of $1 \, \mathrm{dex}$ in production factor below
    the maximum value for each progenitor.
\label{fig:pf}}
\end{figure*}

The peculiar explosion dynamics and progenitor structure close to the
AGB-SN transition have interesting implications for the
nucleosynthesis in this range.  Since the ejected mass from the
tenuous shells (He, C, O) between the core and the hydrogen envelope
is small, primary and secondary hydrostatic and explosive burning
processes in these shells do not contribute significantly in terms of
production factors.  Instead, the yields for
the least massive supernova progenitors are \emph{dominated} by the neutrino-heated material from the gain
region, whose composition, entropy $s$, and electron fraction $Y_e$
are \emph{completely reset} into an equilibrium determined
by neutrino and electron/positron capture reactions
$p (\bar{\nu}_e,e^+) n$,
$n (\nu_e,e^-) p$,
$p (e^-,\nu_e) n$, 
and $n (e^+,\bar{\nu}_e) p$
 in the vicinity of the proto-neutron star
before it is ejected. 
The progenitor's composition and metallicity
therefore do not affect the yields directly. They merely have an
indirect effect via the progenitor structure, e.g., through the
metallicity-dependent location of the AGB-SN transition
\citep{ibeling_13} and the width of the ECSN
channel \citep{poelarends_08}.

These ``innermost'' neutrino-driven ejecta are indeed relevant for the
nucleosynthesis contributions of core-collapse SNe and the
distribution of radionuclides in SN remnants across a wider
mass range of progenitors
\citep{pruet_06,wongwathanarat_16,wanajo_17}.  ECSNe, however, represent the first case for which the complete
nucleosynthesis in the innermost ejecta could be studied in detail
based on self-consistent multi-D explosion models of the
early explosion phase \citep{wanajo_11,wanajo_13a,wanajo_13b} and
1D models of the subsequent neutrino-driven wind
phase \citep{wu_14,pllumbi_15}. The key difficulty here lies in
accurately capturing the evolution of $Y_e$ in the innermost ejecta,
which requires rigorous neutrino transport 
(including even the effects of neutrino flavour conversion
near the proto-neutron star)
to correctly model
differences in the electron neutrino and antineutrino emission 
in contrast to simulations relying on pistons
and thermal bombs 
\citep[e.g.,][]{rauscher_02,limongi_03,tominaga_07,heger_10,nomoto_13,chieffi_13}.
Second, the multi-D explosion dynamics is relevant
as it determines the freeze-out of $Y_e$
at a radius where
$r  \sim
\langle E_\nu\rangle/(m_\mathrm{N} \dot{q}_\nu v_r)$
in terms of the nucleon mass $m_\mathrm{N}$, the mass specific heating rate
$\dot{q}_\nu$, the ejection velocity $v_r$ and the averaged mean
energy $\langle E_\nu \rangle$ of $\nu_e$ and $\bar{\nu}_e$ \citep{qian_96,mueller_16b}.
Finally, simulations need to be evolved sufficiently far to capture
the bulk of the neutrino-heated ejecta and safely demarcate the actual
ejecta from fallback material, which is an impediment for
multi-D explosion models of more massive stars
\citep{wanajo_17} and has so far limited us to
axisymmetric (2D) models as input for nucleosynthesis
calculations. Nevertheless, a comparison of
essentially complete yields for ECSN-like explosions
and the early neutrino-heated ejecta in SNe from more 
massive progenitors is already possible and reveals
pronounced differences.

\section{Neutrino-heated Ejecta in Low-Mass Supernovae}
The key to these differences lies in the development of overturn
driven by the Rayleigh-Taylor instability in the wake of the rapidly
expanding shock between deeper layers of high-entropy neutrino-driven
ejecta and colder ejecta shocked at early times.  Due to high ejection
velocities, $Y_e$ in the rising bubbles freezes out closer to the
proto-neutron star at $Y_e\approx 0.4$ (Fig.~\ref{fig:2d}). Entropies are
modest ($\mathord{\sim} 15 k_\mathrm{b} /\mathrm{nucleon}$) in the most
neutron-rich ejecta and higher in the more slowly expanding ejecta
that have been exposed to neutron heating longer.  \citet{wanajo_11}
showed that for ECSNe the resulting nucleosynthesis is a combination
of freeze-out from $\alpha$-deficient QSE for $Y_e <0.43$ and
$\alpha$-rich QSE for $0.43<Y_e<0.49$ with relatively uniform
production factors between Zn and Zr (Fig.~\ref{fig:pf}, top left).
There is also a significant production of the neutron-rich
radionuclides ${}^{48}\mathrm{Ca}$ \citep{wanajo_13a} and ${}^{60}
\mathrm{Fe}$ \citep{wanajo_13b}. Subsequent work by \citet{wanajo_17}
indicates that the explosion dynamics of low-mass iron core
progenitors is sufficiently extreme to attain the same or at least
similar neutron-rich nucleosynthesis in low-mass iron-core progenitors
with $9.6 M_\odot$ (zero metallicity) and $8.1 M_\odot$ (metallicity
$Z=10^{-4} Z_\odot$) as shown in 
Fig.~\ref{fig:pf} (top row, middle and right).

Slower shock propagation, by contrast, impedes the ejection of
neutron-rich plumes in more massive progenitors with slower shock
propagation, which leads to characteristically different yields from
proton-rich ejecta (bottom row of Fig.~\ref{fig:pf}).  Somewhat
similar yield patterns in the atomic mass range of $A=30\ldots 40$
may, however, also occur for more massive progenitors, as for example
in the $27 M_\odot$ model of \citet{wanajo_17}, where this results
from the ejection of some neutron-rich material in a rather early
explosion and a weak $\nu p$-process in the proto-rich ejecta.

At present, the comparison of nucleosynthesis yields in
ECSN-like explosions and SNe from more massive progenitors
remains beset with many imponderables, including
the impact of neutrino rate uncertainties
and neutrino flavour conversion
(investigated in the context of 1D explosion models
by \citealt{wu_14} and \citealt{pllumbi_15}), 3D
effects on plume ejection \citep{mueller_16b}, the
LESA instability \citep{tamborra_14a} in 3D, resolution
effects and our incomplete understanding of the explosions
of more massive progenitors. 

Nonetheless, self-consistent SN models already
suggest that ECSN-like explosions play a significant role
in a comprehensive picture of chemical evolution, e.g.\
as a source that produces $N=50$ nuclei (Sr, Y, Zr) without
the heavy r-process elements as required by observed
abundance trends \citep{travaglio_04,wanajo_06,qian_08,hansen_13},
or as a source of $^{48} \mathrm{Ca}$ whose origin
remains poorly understood.

\section{Outlook}
With the emerging links between stellar evolution, SN
modelling, and chemogalactic evolution in the case of ECSNe and low-mass iron core supernovae, there is an opportunity
to better constrain the considerable uncertainties that
beset the evolution of the least massive core-collapse SNe
progenitors.  The models of recent years only
constitute the first step in this undertaking. With respect to
SN simulations, some of the salient uncertainties, such as the
lack of 3D first-principle modes, have already been
mentioned. Moreover, SN simulations need to
scan the low-mass end of the spectrum more thoroughly
given the multiplicity of subtly different stellar evolution
channels in a small mass window.

In addition to encouraging progress in confronting models of ECSN-like
explosions with nucleosynthesis constraints, it remains imperative to
better understand the nature of the diverse low-energy transients that
are \emph{prima facie} suggestive of ECSN-like explosion dynamics with
low explosion energies and small nickel masses. Noteworthy results on
the light curve of SN~1054 (the Crab supernova) and various Type~IIn-P
SNe have been obtained recently \citep{tominaga_13,moriya_14}
to strengthen the suggested link between ECSNe and these events
\citep[e.g.,][]{smith_13}. Progenitor detections (or upper limits on progenitor
brightness in lieu of a positive detection) are also helping to match
low-energy transients to the various SN channels close to the
AGB-SN mass transition (SN~2005cs: \citealt{eldridge_07}s, SN~2008S: \citealt{botticella_09}).
Distinguishing these channels by means of observations remains a
challenge, however, and a definitive ``smoking gun'' for the classical
ECSN channel of SAGB stars is still missing. A better connection of
stellar evolution and explosion models to the signatures of the
photospheric and nebular phase (see A.~Jerkstrand, these proceedings)
is still needed to accomplish this.

\begin{acknowledgements}
We acknowledge support by ARC grants DE150101145 (BM) and FT120100363
(AH), STFC grant ST/P000312/1 
(BM), 
by the RIKEN iTHES Project,
the JSPS Grants-in-Aid for Scientific Research (26400232, 26400237)
(SW), the Deutsche Forschungsgemeinschaft (EXC~153), by the ERC grant
ERC-AdG No.~341157-COCO2CASA (TJ), and NSF Grant
No.~PHY-1430152 through JINA-CEE (AH, BM). Supercomputer time at the National Computational
Infrastructure (NCI), the Pawsey Supercomputing Centre, the Max Planck
Computing and Data Facility, the Minnesota Supercomputing Institute,
and the Dirac Data Centric system (Durham) is acknowledged.

\end{acknowledgements}

\bibliography{paper}

\end{document}